\documentclass[12pt]{article}
\newcommand{\thetav}{\mbox{\boldmath$\theta$}}
\begin{document}

%%%%%%%%%%%%%%%%%%%%
\title{{\bf{\Large  On the role of twisted statistics in the noncommutative degenerate electron gas}}}
%%%%%%%%%%%%%%%%%%%%
\author{
{\bf {\normalsize Subrata Khan}$^{a}$\thanks{skhan@bose.res.in}}, {\bf {\normalsize Biswajit Chakraborty}$^{a,b}$\thanks{biswajit@bose.res.in}}\\ {\bf {\normalsize Frederik G. Scholtz}$^{a,b}$\thanks{fgs@sun.ac.za}}\\
$^a$ {\normalsize S.~N.~Bose National Centre for Basic Sciences,}\\{\normalsize JD Block, Sector III, Salt Lake, Kolkata-700098, India}\\[0.3cm]
$^b$ {\normalsize Institute of Theoretical Physics, University of Stellenbosch,}\\ {\normalsize Stellenbosch 7600, South Africa}
}
\date{}

\maketitle

%%%%%%%%%%%%%%%%
\begin{abstract}
%%%%%%%%%%%%%%%%

We consider the problem of a degenerate electron gas in the background of a uniformly distributed positive
charge, ensuring overall neutrality of the system, in the presence of non-commutativity. In contrast to previous calculations \cite{bem} that did not include twisted statistics, we find corrections to the ground state energy already at first order in perturbation theory when the twisted statistics is taken into account.  These corrections arise since the interaction energy is sensitive to two particle correlations, which are modified for twisted anti-commutation relations.
\\[0.3cm]
\\[0.3cm]

\end{abstract}

\section{Introduction}
\label{int}

The study of non-commutative (NC) geometry and its implications have gained considerable importance in recent times as
these studies are motivated from string theory and certain condensed matter systems like the quantum Hall effect.  Despite this, the physical consequences of non-commutativity remains unclear. In free space (no boundaries) and in the absence of interactions, non-commutativity seems to have no observable physical consequences \cite{gov,wess}.  On the other hand, in \cite{bem} it was found that non-commutativity does effect the ground state energy of a degenerate electron gas to second order in perturbation theory.  However, in this computation the role of twisted statistics was not taken into account. As is well known by now it is necessary to twist the anti-commutation relations in order to restore the Galilean or, more generally, Poincare invariance \cite{chai,bal}. Here we revisit this calculation taking also due care of the twisted statistics.  One expects that the twisted statistics will also have an effect on the ground state energy as it changes the two particle correlations \cite{chak} and it is well known that the two particle interaction energy is sensitive to the two particle correlations.  Thus there are two possible ways in which non-commutativity may have a physical effect. The first is due to the non-commutative nature of space per se and the second the modification of two particle correlations due to the twisted statistics required to restore Poincare invariance.   

Here we demonstrate this in the specific setting of \cite{bem}, namely, a non-relativistic degenerate electron gas in the presence of a uniform neutralising background charge, interacting through a screened coulomb potential. One of the motivations for studying this particular system is that this is the typical setting encountered in astro physical or quantum Hall systems.  

It is generally believed that the non-commutative parameter $\theta^{\mu\nu}$ is of the order of the Planck area.
It, therefore, seems very unlikely that any experimental signature of its presence can be detected by any terrestial observation, unless one finds a way of amplifying it by several orders of magnitude. One of the motivations for this paper is to study whether this is at all feasible by considering a non-relativistic degenerate electron gas, which presumably can contain few orders higher than Avogadro's number ($\sim 10^{23}$) of particles, so that the effects of non-commutativity may be amplified. 

The plan of the paper is as follows. We first introduce the basic notations and conventions of NC geometry in Section \ref{sec2}. In Section \ref{sec3} we provide a brief review on the origin of twisted (anti)-commutation relations. In Section \ref{sec4} we consider the effect of the non-commutativity by computing the energy-shift, arising from the screened inter-particle Coulomb potential, to first order in perturbation theory. Here we use the technology developed by Fetter and Walecka \cite{fet} for the same commutative problem ($\theta = 0$).Finally
we conclude in Section \ref{con}.  

%---------------------------------------------------------------------------
\section{NC geometry}
\label{sec2}
%---------------------------------------------------------------------------
    
To fix our notation and conventions, we briefly recall the essentials of a NC geometry. The canonical NC structure is given by the following operator valued space-time coordinates,
\begin{eqnarray}
[x^{\mu}_{op}, x^{\nu}_{op}] = i\theta^{\mu\nu}\label{comm}
\end{eqnarray}
Instead of working with functions of these operator-valued coordinates, one can alternatively work with functions of c-numbered coordinates provided one composes the functions (taken to belong to the Schwartz Class of functions $\it {i.e.}$ whose derivatives of all order, both in configuration and momentum space, vanishes asymtotically) through the Moyal star ($*$) product defined as\cite{szabo}
\begin{eqnarray}
\alpha*_{\theta}\beta(x) = [\alpha \exp(\frac{i}{2}\buildrel{\leftarrow}\over{\partial}_{\mu}\theta^{\mu\nu} \buildrel{\rightarrow}\over{\partial}_{\nu}) \beta](x)\label{star}\\
\theta^{\mu\nu} = -\theta^{\nu\mu} \in R, \quad x = (x^{0}, x^{1},....,x^{d}).
\end{eqnarray}
With this, the Moyal bracket $[x^{\mu},x^{\nu}]_{*} \equiv (x^{\mu}*x^{\nu} - x^{\nu}*x^{\mu}) = i\theta^{\mu \nu}$ is isomorphic to the corresponding commutator (\ref{comm}) involving $x^{\mu}_{op}$. In this paper, we consider only the problem of spatial non-commutativity and therefore set $\theta^{0i} = 0$.
The Poincare group $\cal P$ or the diffeomorphism group $\cal D$, which acts on the NC space-time $R^{d+1}$, defines a natural action on smooth functions $\alpha \in {\cal C}^{\infty}(R^{d+1})$ as
\begin{eqnarray}
(g\alpha)(x) = \alpha(g^{-1}x),
\end{eqnarray}
for g $\in \cal P$ or $\in \cal D$.  However, in general
\begin{eqnarray}
(g\alpha)*_{\theta} (g\beta) \neq g(\alpha*_{\theta}\beta),
\end{eqnarray}
showing that the action of the group $\cal P$ or $\cal D$ is not an automorphism of the Algebra ${\cal A}_{\theta} (R^{d+1})$, unless one considers the translational sub-group.
The Poincare symmetry can, however, be restored by the twisted implementation of the Lorentz group, as has been
shown recently in the literature \cite{chai,bal}. This in turn implies that the permutation group, used for defining bosons or
fermions, gets twisted as well in order to maintain the statistics operator as a super-selected observable. One thus ends up
defining twisted fermionic or bosonic fields, the Fourier modes of which are subjected to the following (anti)commutation relations:
\begin{eqnarray}
\label{twcr1}
\hat{\tilde{a}}_{{\bf k}_{1}\lambda_{1}}\hat{\tilde{a}}_{{\bf k}_{2}\lambda_{2}} = \eta e^{2i{\bf k}_{1}\wedge{\bf k}_{2}}\hat{\tilde{a}}_{{\bf k}_{2}\lambda_{2}}\hat{\tilde{a}}_{{\bf k}_{1}\lambda_{1}},
\end{eqnarray}
\begin{eqnarray}
\label{twcr2}
\hat{\tilde{a}}_{{\bf k}_{1}\lambda_{1}}^{\dag}\hat{\tilde{a}}_{{\bf k}_{2}\lambda_{2}}^{\dag} = \eta e^{2i{\bf k}_{1}\wedge{\bf k}_{2}}\hat{\tilde{a}}_{{\bf k}_{2}\lambda_{2}}^{\dag}\hat{\tilde{a}}_{{\bf k}_{1}\lambda_{1}}^{\dag},
\end{eqnarray}
\begin{eqnarray}
\label{twcr3}
\hat{\tilde{a}}^{\dag}_{{\bf k}_{1}\lambda_{1}}\hat{\tilde{a}}_{{\bf k}_{2}\lambda_{2}} = \frac{1}{\eta}(\hat{\tilde{a}}_{{\bf k}_{2}\lambda_{2}}\hat{\tilde{a}}^{\dag}_{{\bf k}_{1}\lambda_{1}}-(2\pi)^{3}\delta({\bf k}_{2} - {\bf k}_{1})\delta_{\lambda_{1}\lambda_{2}})e^{2i{\bf k}_{2}\wedge{\bf k}_{1}},
\end{eqnarray}
where ${\hbar \bf k}$ represents the spatial components of momentum and the $\{\lambda\}$'s represent the spin degree of freedom. Incidentally, this structure of twisted (anti)-commutation relation can also be obtained in the non-relativistic domain, if one considers the twisted action of the Galileo group \cite{chak}. As has been shown in the paper just mentioned, one has to just restore the symmetry corresponding to spatial rotation (SO(3)) here, as the Galileo boost generator does not get affected by the twist. Also the wedge symbol ($\wedge$) inserted between ${\bf k}_{1}$ and ${\bf k}_{2}$ denotes
\begin{eqnarray}
{\bf k}_{1}\wedge{\bf k}_{2} = \frac{1}{2}k_{1i}\theta^{ij}k_{2j}.
\end{eqnarray}
Finally, $\eta = \pm 1$ as dictated by Boson or Fermion statistics. As we are interested in fermionic systems $\eta = -1$. These operators go over to usual Bose/Fermi ladder operators in the limit $\theta \rightarrow 0$: $\hat{\tilde{a}}\rightarrow \hat{a}$. In fact, it has been shown in \cite{chai} that $\hat{\tilde{a}}$ and $\hat{a}$ are related as $\hat{\tilde{a}}=e^{(\frac{i}{2}p_{\mu}\theta^{\mu\nu}P_{\nu})}\hat{a}$.

%---------------------------------------------------------------------------
\section{A brief review of twisted (anti)-commutation relation}
\label{sec3}
%---------------------------------------------------------------------------
It has been shown earlier that in a non-relativistic system, the presence of spatial non-commutativity given by $\theta^{ij}$ spoils the SO(3) rotational symmetry. This can be seen easily from the fact that the vector ${\thetav} = \{ \theta_{i} \}$, dual to the second rank anti-symmetric tensor $\theta^{ij}$
\begin{eqnarray}
\theta_{i}= \frac{1}{2}\epsilon_{ijk}\theta^{jk},
\end{eqnarray}
fixes a direction in the 3-dimensional space.  Thus the only surviving symmetry is the SO(2) rotation around the ${\thetav}$ direction. However, applying the Drinfeld twist to the Hopf algebra, the entire SO(3) symmetry can be restored. This is done by considering a twisted co-product, $\Delta_{\theta}(M_{ij})$, corresponding to SO(3)-generators $M_{ij}$, with action defined as
\begin{eqnarray}
\Delta_{\theta}(M_{ij}) = \it{F}_{\theta}^{-1}\Delta_{0}(M_{ij})\it{F}_{\theta},
\end{eqnarray}
where
\begin{eqnarray}
\Delta_{0}(M_{ij})= M_{ij}\bigotimes 1 + 1\bigotimes M_{ij}
\end{eqnarray}
is the co-product in the commutative case $(\theta =0)$ and $\it{F}_{\theta} = e^{-\frac{i}{2}\theta^{ij}P_{i}\bigotimes P_{j}}$ is the twist operator. The corresponding co-product for the rotation group SO(3) in the commutative and non-commutative cases are 
\begin{eqnarray}
\Delta_{0}(g)= g \bigotimes g , \quad \Delta_{\theta}(g) = \it{F}_{\theta}^{-1}\Delta_{0}(g)\it{F}_{\theta}.
\end{eqnarray}
Note that the star-product, defined in (\ref{star}), can now be written alternatively as
\begin{eqnarray}
(\alpha * \beta)(x)= m_{\theta}(\alpha \bigotimes \beta) = m_{0}(\it{F}_{\theta} \alpha \bigotimes \beta)
\end{eqnarray}
where $m_{0}$ represents the composition of a pair of functions in the commutative $(\theta =0)$ case, which is nothing but point-wise multiplication of these functions,i.e, $\alpha(x)\beta(x)$.

It can be easily seen at this stage that the usual projection operator for a two-particle system, $P_{0}=\frac{1}{2}(1 \pm \tau_{0})$, involving the flip map $\tau_{0}(\alpha \bigotimes \beta)= \beta \bigotimes \alpha$, and which projects onto the symmetric (anti-symmetric) sub-space describing bosonic (fermionic) statistics, does no longer commute with $\Delta_{\theta}$: $[\Delta_{\theta},P_{0}] \neq 0$. This implies that one must twist the flip map as 
\begin{eqnarray}
\label{twist}
\tau_{\theta} = \it{F}_{\theta}^{-1}\tau_{0}\it{F}_{\theta} = \it{F}_{\theta}^{-2}\tau_{0},
\end{eqnarray}
so that the corresponding projection operator $P_{\theta}=\frac{1}{2}(1\pm \tau_{\theta}) = \it{F}_{\theta}^{-1}P_{0}\it{F}_{\theta}$ commutes with $\Delta_{\theta}$
\begin{eqnarray}
[\Delta_{\theta}, P_{\theta}] = 0.
\end{eqnarray}
This ensures that the new statistics, $\it{i.e.}$ the flip operator $\tau_{\theta}$, remains super-selected, $\it{i.e.}$ it commutes with all Galilean generators, and defines twisted bosons or fermions. We now provide a heuristic argument how the algebra involving the corresponding creation and anihilation operators also get modified.

To begin with, let us apply the twisted projection operator $P_{\theta}$ on the tensor product of two momentum eigen- states $|k>\bigotimes |l>$ and make use of (\ref{twist}):
\begin{eqnarray}
P_{\theta}(|k>\bigotimes |l>)=\frac{1}{2}(|k>\bigotimes |l>+ \eta \it{F}_{\theta}^{-2}|l>\bigotimes |k>).
\end{eqnarray}
Up to an overall phase, this can be re-written in a symmetric or anti-symmetric form as in the $\theta = 0$ case
\begin{eqnarray}
P_{\theta}(|k>\bigotimes |l>) = e^{ik\wedge l}[\frac{1}{2}(|k,l>>+\eta |l,k>>)], \nonumber \\
|k,l>> \equiv e^{-ik\wedge l}|k>\bigotimes |l>.
\end{eqnarray}
Now identifying $\hat{\tilde{a}}_{k}^{\dag}\hat{\tilde{a}}_{l}^{\dag} |0> = P_{\theta}(|k>\bigotimes |l>)$ it easily follows that $\hat{\tilde{a}}_{k}^{\dag}\hat{\tilde{a}}_{l}^{\dag} = \eta e^{2ik\wedge l}\hat{\tilde{a}}_{l}^{\dag}\hat{\tilde{a}}_{k}^{\dag}$.
The other phase $e^{-2ik\wedge l}$, occuring in $\hat{\tilde{a}}_{k}^{\dag}\hat{\tilde{a}}_{l}$ of eq. (\ref{twcr3}), can be easily understood from the fact that the anihilation operator $\hat{\tilde{a}}_{l}$ is associated with a momentum $(-l)$, in contrast to the operator $\hat{\tilde{a}}_{l}^{\dag}$, for which the associated momentum is $(+l)$.

%---------------------------------------------------------------------------
\section{Non-commutative degenerate electron gas}
\label{sec4}
%---------------------------------------------------------------------------
To begin with, the Hamiltonian operator for the electronic system in the second-quantized formulation can be written as
\begin{eqnarray}
\label{ham}
\hat{H}_{el} = \int{d^{3}x\hat{\psi}^{\dag}({\bf x})*\hat{T}\hat{\psi}({\bf x})} + \frac{1}{2}\int\int{d^{3}xd^{3}y\hat{\psi}^{\dag}({\bf x})*\hat{\psi}^{\dag}({\bf y})*V({\bf x},{\bf y})*\hat{\psi}({\bf y})*\hat{\psi}({\bf x})}.
\end{eqnarray}
Before we explain the terms and notations, first note that in order to get a well-defined thermodynamic limit, we introduce a box of volume $V=L^{3}$, containing $N$ particles, as a regulator so that when $V \rightarrow \infty$ and $N \rightarrow \infty$, the density $(N/V)$ is held fixed.
This implies that the field operators
\begin{eqnarray}
\hat{\psi}({\bf x}) = \sum_{{\bf k},\lambda}\psi_{{\bf k}\lambda}({\bf x})\hat{\tilde{a}}_{{\bf k}\lambda},
\end{eqnarray}
occuring in (\ref{ham}) act on the Fock-space of states obtained by superposing anihilation operators in terms of the single particle wavefunctions
\begin{eqnarray}
\label{boxwf}
\psi_{{\bf k}\lambda}({\bf x}) = \frac{1}{\sqrt V}e^{i{\bf k}\cdot{\bf x}}\eta_{\lambda}
\end{eqnarray}
with ${\bf k}$ taking the discrete set of values ${\bf k} = \frac{2\pi}{L} {\bf n}$ with ${\bf n}$ representing a triplet of positive or negative integers, arising from the imposition of periodic boundary conditions in the box, 
and $\eta_{\lambda}$ stands for two spin functions, $\eta_{\uparrow} = \pmatrix{1\cr 0},\quad \eta_{\downarrow} = \pmatrix{0\cr 1}$.  Furthermore, the arguments ${\bf x}$ are now the usual c-numbered coordinates, so that they compose through $*$-products. In this form the action is hermitian, as follows from the fact that $(f*g)^{\dag}=g^{\dag}*f^{\dag}$ \cite{szabo}. Here we would like to clarify some conceptual issues. These are related to the concept of a box in NC space. First of all, at the level of operators $x^{\mu}_{op}$ satisfying (\ref{comm}) the concept of a well delineated boundary does not make sense, as the uncertainty relation $|\Delta x^{i}||\Delta x^{j}| \geq \frac{|\theta^{ij}|}{2}$ following from (\ref{comm}) prevents this. However, at the level of commuting $x^{i}$'s, when the Moyal star-product is being used, one can presumably introduce a boundary in the Moyal plane if the "location" is interpreted in the sense of "expectation value". Secondly, for particles confined within a box, a typical single-particle wave function has a support only within the box, where for a momentum eigenstate this can be taken to be in the plane-wave form (\ref{boxwf}). However, these wave functions with a compact support do not belong to the Schwartz Class, as mentioned earlier. The best that we can do at this stage is to extend the domain of definition of the function from the box to the entire three-dimensional space through periodicity and then compose any pair of such plane-wave functions through the star-product. The resultant function, which incidentally is also of the same form (\ref{boxwf}) up to a phase factor, is then restricted within the box. One can expect that this prescription will yield the desired result in the thermodynamic limit $L \rightarrow \infty$.

The kinetic and interaction energy operators are denoted by $\hat{T}({\bf x})$ and $V({\bf x}, {\bf y})$, respectively:
\begin{eqnarray}
\hat{T} = \frac{1}{2m}\hat{{\bf p}}^{2} = -\frac{\hbar^{2}}{2m}{\bf \nabla}^{2},
\end{eqnarray}
\begin{eqnarray}
V({\bf x}, {\bf y}) = \frac{e^{2}}{2}\frac{e^{-\mu|{\bf x}-{\bf x}|}}{|{\bf x}-{\bf x}|}.
\end{eqnarray}
This has to be augmented by the Hamiltonian of the positive inert background having the particle density $n({\bf x})$
\begin{eqnarray}
H_{b} = \frac{e^{2}}{2}\int{d^{3}xd^{3}x'\frac{n(x)n(x')e^{-\mu|{\bf x} - {\bf x}'|}}{|{\bf x} - {\bf x'}|}}
\end{eqnarray}
and the Hamiltonian
\begin{eqnarray}
H_{el-b} = -e^{2}\sum_{i=1}^{N}\int{d^{3}x\frac{n(x)e^{-\mu|{\bf x} - {\bf x}_{i}|}} {|{\bf x} - {\bf x}_{i}|}},
\end{eqnarray}
representing the energy between the electrons and positive background. With this the total Hamiltonian for the system becomes
\begin{eqnarray}
\label{totham}
H = H_{el} + H_{b} + H_{el-b}.
\end{eqnarray}

Here we have introduced a screened Coulomb potential through an exponentially damping factor having a regulator $\mu$ of dimension $[L]^{-1}$. This renders the integrals appearing in $H_{b}$ and $H_{el-b}$ finite in the thermodynamic limit. It should, however, be kept in mind that the limit $\mu \rightarrow 0$ should be taken after the $L \rightarrow \infty$ limit, so that one can ensure $\mu^{-1}\ll L$ at each step of the computation. This also facilitates the shifting of origin of integration, as dictated by convenience, as the surface terms are negligibly small in this limit.

Taking the distribution to be uniform $n({\bf x}) = N/V$, we can now compute the pair of non-dynamical (inert) terms
\begin{eqnarray}
H_{b} = \frac{1}{2}e^{2}\frac{N^{2}}{V}\frac{4\pi}{\mu^{2}},
\end{eqnarray}
\begin{eqnarray}
H_{el-b} = -e^{2}\frac{N^{2}}{V}\frac{4\pi}{\mu^{2}},
\end{eqnarray}
where we have made use of the translational invariance. As expected, these are c-number terms.
The total Hamiltonian (\ref{totham}) thus reduces to
\begin{eqnarray}
\label{totham1}
H = -\frac{1}{2}e^{2}\frac{N^{2}}{V}\frac{4\pi}{\mu^{2}} + H_{el}
\end{eqnarray}
with all the interesting physical effects being buried in $H_{el}$.

We therefore turn our attention to $H_{el}$. To begin with, let us consider the kinetic term in (\ref{ham}) first. This can be simplified as,
\begin{eqnarray}
\int{d^{3}x\hat{\psi}^{\dag}({\bf x})*\hat{T}\hat{\psi}({\bf x})} = \sum_{{\bf k}\lambda {\bf k}^{\prime} \lambda^{\prime}}\frac{\hbar^{2}{\bf k}^{2}}{2m}\hat{\tilde{a}}_{{\bf k}^{\prime}\lambda^{\prime}}^{\dag}\hat{\tilde{a}}_{{\bf k}\lambda}\int{d^{3}x\psi_{{\bf k}^{\prime}\lambda^{\prime}}^{\dag}({\bf x})*\psi_{{\bf k}\lambda}({\bf x})}.
\end{eqnarray}
Using the definition of the $*$-product given in (\ref{star}), we can easily see that this brings in an exponential factor $e^{-i{\bf k}^{\prime}\wedge {\bf k}}$, as the $*$-operator is sandwitched between a pair of plane-wave states of the form (\ref{boxwf}). However, the integration yields $\delta_{{\bf k}{\bf k}^{\prime}}\delta_{\lambda \lambda^{\prime}}$, thereby reducing this $\theta$-dependent exponential factor to identity. As far as this kinetic term is concerned, this therefore yields the same result as in the commutative ($\theta = 0$) case:
\begin{eqnarray}
\int{d^{3}x\hat{\psi}^{\dag}({\bf x})*\hat{T}\hat{\psi}({\bf x})} = \sum_{{\bf k}\lambda}\frac{\hbar^{2}k^{2}}{2m}\hat{\tilde{a}}_{{\bf k}\lambda}^{\dag}\hat{\tilde{a}}_{{\bf k}\lambda},
\end{eqnarray}
which can be interpreted as the kinetic energy of each mode multiplied by the corresponding number operator. Almost the same thing also happens for the potential energy term as all $\theta$-dependent exponential factors coming from the Moyal star-product reduces to the identity here as well. Seemingly, the non-commutative structure of space, encoded in the Moyal star-product, plays no role, at least to first order in perturbation theory.  This was also found in \cite{bem}.  There the effects of non-commutativity only showed up in second order, which is not unexpected as the exchange correlations of a non-commutative theory will generacilly be different from those of a commutative theory.  However, as we proceed to show now, a $\theta$-dependence, stemming from the twisted anti-commutation relation, survives to give a NC modification to the ground-state energy already to first order in perturbation theory.  

To this end consider the potential term in (\ref{ham}). First of all this requires a proper interpretation. Given that ${\bf x}$ and ${\bf y}$ represents the position co-ordinates of two distinct particles in a pair, we must have $[x^{\mu},y^{\nu}]_{*} = 0$. Consequently, only functions involving coordinates of a single-particle  will compose through the $*$-product. This implies that the appropriate way of writing this potential term is
\begin{eqnarray}
\int{d^{3}x\hat{\psi}^{\dag}({\bf x})*(\int{d^{3}y}\hat{\psi}^{\dag}({\bf y})*V({\bf x},{\bf y})*\hat{\psi}({\bf y}))*\hat{\psi}({\bf x})},
\end{eqnarray}
where the ${\bf y}$-integral is performed first, after composing the three functions of ${\bf y}$, with ${\bf x}$ being held fixed, to yield a function of ${\bf x}$. With this interpretation, we can write 
\begin{eqnarray}
\int\int{d^{3}xd^{3}y\hat{\psi}^{\dag}({\bf x})*\hat{\psi}^{\dag}({\bf y})*V({\bf x},{\bf y})*\hat{\psi}({\bf y})*\hat{\psi}({\bf x})} = \frac{1}{2}\sum_{\{k\}\{\lambda\}}\hat{\tilde{a}}_{{\bf k}_{1}\lambda_{1}}^{\dag}\hat{\tilde{a}}_{{\bf k}_{2}\lambda_{2}}^{\dag}\hat{\tilde{a}}_{{\bf k}_{3}\lambda_{3}}\hat{\tilde{a}}_{{\bf k}_{4}\lambda_{4}}\times \nonumber \\
<{\bf k}_{1}\lambda_{1}{\bf k}_{2}\lambda_{2}|{\it V}|{\bf k}_{3}\lambda_{3}{\bf k}_{4}\lambda_{4}>,
\end{eqnarray}
where the matrix element is given by
\begin{eqnarray}
<{\bf k}_{1}\lambda_{1}{\bf k}_{2}\lambda_{2}|{\it V}|{\bf k}_{3}\lambda_{3}{\bf k}_{4}\lambda_{4}> = \frac{e^{2}}{2V^{2}}\int{d^{3}x e^{-i{\bf k}_{1}.{\bf x}}\eta_{\lambda_{1}}(1)^{\dag}*(\int{d^{3}ye^{-i{\bf k}_{2}.{\bf y}}\eta_{\lambda_{2}}(2)^{\dag}*}} \nonumber \\ 
\frac{e^{-\mu|{\bf x} - {\bf y}|}} {|{\bf x} - {\bf y}|}*}e^{i{\bf k}_{4}.{\bf y}}\eta_{\lambda_{4}}(2))*e^{i{\bf k}_{3}.{\bf x}}\eta_{\lambda_{3}(1).
\end{eqnarray}
Making use of the identity $\frac{e^{-\mu|{\bf x} - {\bf y}|}} {|{\bf x} - {\bf y}|} = \frac{4\pi}{(2\pi)^{3}}\int{\frac{d^{3}k}{\mu^{2}+k^{2}}e^{i{\bf k}.({\bf x}-{\bf y})}}$ {\footnote {This can easily be seen to follow by noting that the screened Coulomb potential is a Green's function of the Laplacian augmented by a mass term: $(-\nabla^{2} + \mu^{2}) (\frac{e^{-\mu|{\bf x}|}} {|{\bf x}|}) = 4\pi \delta^{3}({\bf x})$}}we can simplify this matrix element to get
\begin{eqnarray}
<{\bf k}_{1}\lambda_{1}{\bf k}_{2}\lambda_{2}|{\it V}|{\bf k}_{3}\lambda_{3}{\bf k}_{4}\lambda_{4}> = \frac{4\pi}{(2\pi)^{3}}\frac{e^{2}}{2}\int{\frac{d^{3}k}{\mu^{2}+k^{2}}e^{-i{\bf k}_{1}\wedge{\bf k}_{3}}e^{-i{\bf k}_{2}\wedge{\bf k}_{4}}\delta_{\lambda_{1}\lambda_{3}}\delta_{\lambda_{2}\lambda_{4}}} \times \nonumber \\ \delta_{({\bf k}_{1}+{\bf k}_{2}),({\bf k}_{3}+{\bf k}_{4})},
\end{eqnarray}
where ${\bf k}$ gets restricted to ${\bf k} = {\bf k}_{1} - {\bf k}_{3} = {\bf k}_{4} - {\bf k}_{2}$, so that ${\bf k}$, occuring in the Fourier transform of the Screened Coulomb potential, can be identified with the momentum transfer. On the other hand, the Kronecker delta involving momenta levels enforces momentum conservation. We therefore convert integration over '$k$' to a sum by the standard replacement $\int{d^{3}k} \rightarrow (\frac{2\pi}{L})^{3} \sum_{{\bf k}}$, so that the parity with other momentum variables can be restored. We thus get,
\begin{eqnarray}
<{\bf k}_{1}\lambda_{1}{\bf k}_{2}\lambda_{2}|{\it V}|{\bf k}_{3}\lambda_{3}{\bf k}_{4}\lambda_{4}> = \frac{2\pi e^{2}}{V}\sum_{\bf k} \frac{1}{\mu^{2}+k^{2}}e^{-i{\bf k}_{1}\wedge{\bf k}_{3}}e^{-i{\bf k}_{2}\wedge{\bf k}_{4}}\delta_{\lambda_{1}\lambda_{3}}\delta_{\lambda_{2}\lambda_{4}}\delta_{({\bf k}_{1}+{\bf k}_{2}),({\bf k}_{3}+{\bf k}_{4})}.
\end{eqnarray}
The potential energy operator becomes
\begin{eqnarray}
\frac{2\pi e^{2}}{V}\sum_{\{\bf k\}\{\lambda\}}\frac{1}{\mu^{2}+k^{2}}e^{-i{\bf k}_{1}\wedge{\bf k}_{3}}e^{-i{\bf k}_{2}\wedge{\bf k}_{4}}\delta_{\lambda_{1}\lambda_{3}}\delta_{\lambda_{2}\lambda_{4}}\delta_{({\bf k}_{1}+{\bf k}_{2}),({\bf k}_{3}+{\bf k}_{4})}\tilde{a}_{{\bf k}_{1}\lambda_{1}}^{\dag}\tilde{a}_{{\bf k}_{2}\lambda_{2}}^{\dag}\tilde{a}_{{\bf k}_{4}\lambda_{4}}\tilde{a}_{{\bf k}_{3}\lambda_{3}},
\end{eqnarray}
where $\{{\bf k}\}$ represents the set $({\bf k}, {\bf k}_{1}, {\bf k}_{2}, {\bf k}_{3}, {\bf k}_{4})$ on which the summation has to be performed. 
The total Hamiltonian can now be written as
\begin{eqnarray}
\label{totham2}
\hat H = -\frac{1}{2}e^{2}\frac{N^{2}}{V}\frac{4\pi}{\mu^{2}} + \sum_{{\bf k}\lambda}\frac{\hbar^{2}k^{2}}{2m}\tilde{a}_{{\bf k}\lambda}^{\dag}\tilde{a}_{{\bf k}\lambda} + \frac{2\pi e^{2}}{V}\sum_{\{\bf k\}\{\lambda\}}\frac{1}{\mu^{2}+k^{2}}e^{-i{\bf k}_{1}\wedge{\bf k}_{3}}e^{-i{\bf k}_{2}\wedge{\bf k}_{4}}\delta_{\lambda_{1}\lambda_{3}}\delta_{\lambda_{2}\lambda_{4}}\times \nonumber \\ \delta_{({\bf k}_{1}+{\bf k}_{2}),({\bf k}_{3}+{\bf k}_{4})}\tilde{a}_{{\bf k}_{1}\lambda_{1}}^{\dag}\tilde{a}_{{\bf k}_{2}\lambda_{2}}^{\dag}\tilde{a}_{{\bf k}_{4}\lambda_{4}}\tilde{a}_{{\bf k}_{3}\lambda_{3}}.
\end{eqnarray}

The electrical neutrality of the system makes it possible to eliminate $\mu$ from the Hamiltonian. The last term of the Eq. (\ref{totham2}) can now be re-cast using the momentum transfer ${\bf k}$ as
\begin{eqnarray}
\frac{2\pi e^{2}}{V}\sum_{[{\bf k}{\bf k}_{3}{\bf k}_{4}\lambda_{1}\lambda_{2}]}\frac{1}{\mu^{2}+k^{2}}e^{-i{\bf k}\wedge{\bf k}_{3}}e^{i{\bf k}\wedge{\bf k}_{4}}\tilde{a}_{{\bf k}+{\bf k}_{3},\lambda_{1}}^{\dag}\tilde{a}_{{\bf k}_{4}-{\bf k},\lambda_{2}}^{\dag}\tilde{a}_{{\bf k}_{4}\lambda_{2}}\tilde{a}_{{\bf k}_{3}\lambda_{1}},\label{lastterm}
\end{eqnarray}
where the two spin summation have been evaluated with the Kronecker deltas. At this stage, it is convenient to seperate the above expression into two terms, refering to ${\bf k} \neq 0$ and ${\bf k} = 0$, respectively,
\begin{eqnarray} 
\frac{2\pi e^{2}}{V}\sum_{[{\bf k}{\bf k}_{3}{\bf k}_{4}\lambda_{1}\lambda_{2}]}'\frac{1}{\mu^{2}+k^{2}}e^{-i{\bf k}\wedge{\bf k}_{3}}e^{i{\bf k}\wedge{\bf k}_{4}}\tilde{a}_{{\bf k}+{\bf k}_{3},\lambda_{1}}^{\dag}\tilde{a}_{{\bf k}_{4}-{\bf k},\lambda_{2}}^{\dag}\tilde{a}_{{\bf k}_{4}\lambda_{2}}\tilde{a}_{{\bf k}_{3}\lambda_{1}} + \nonumber \\ \frac{2\pi e^{2}}{V}\sum_{[{\bf k}_{3}{\bf k}_{4}\lambda_{1}\lambda_{2}]}\frac{1}{\mu^{2}}\tilde{a}_{{\bf k}_{3}\lambda_{1}}^{\dag}\tilde{a}_{{\bf k}_{4}\lambda_{2}}^{\dag}\tilde{a}_{{\bf k}_{4}\lambda_{2}}\tilde{a}_{{\bf k}_{3}\lambda_{1}}
\end{eqnarray} 
where the prime on the first summation means the ${\bf k} = 0$ term is omitted. The second term may be rewritten with the discrete version of twisted anticommutation relation given by (\ref{twcr1}), (\ref{twcr3}). In this discrete version, (\ref{twcr1}) remains unaffected, while one has to just replace $(2\pi)^{3}\delta^{3}({\bf k}_{1}- {\bf k}_{2}) \rightarrow \delta_{{\bf k}_{1}{\bf k}_{2}}$ in (\ref{twcr3}). Using this the second term takes the form,
\begin{eqnarray}
\frac{2\pi e^{2}}{V}\sum_{[{\bf k}_{3}{\bf k}_{4}\lambda_{1}\lambda_{2}]}\frac{1}{\mu^{2}}\frac{1}{\eta}(\tilde{a}^{\dag}_{{\bf k}_{3}\lambda_{1}}\tilde{a}_{{\bf k}_{3}\lambda_{1}}\tilde{a}^{\dag}_{{\bf k}_{4}\lambda_{2}}\tilde{a}_{{\bf k}_{4}\lambda_{2}} - \tilde{a}^{\dag}_{{\bf k}_{3}\lambda_{1}}\delta_{{\bf k}_{3}{\bf k}_{4}}\delta_{\lambda_{1}\lambda_{2}}\tilde{a}_{{\bf k}_{4}\lambda_{2}})\eta e^{2i{\bf k}_{3}\wedge{\bf k}_{4}}e^{2i{\bf k}_{4}\wedge{\bf k}_{3}} \nonumber \\ = \frac{2\pi e^{2}}{V}\frac{4\pi}{\mu^{2}}(\hat{N}^{2}-\hat{N}),
\end{eqnarray}
where 
\begin{eqnarray}
\hat N = \int d^{3}x \hat n (x) = \sum_{r} \hat{\tilde{a}}_{r}^{\dag} \hat{\tilde{a}}_{r} = \sum_{r} \hat{a}_{r}^{\dag}\hat{a}_{r} = \sum_{r} \hat n_{r} = \int d^{3}x \hat{\psi^{\dag}}(x)\hat{\psi}(x)
\end{eqnarray}
represents the number operator. Here too the effect of non-commutativity disappears.
{\footnote {As has been mentioned earlier it was shown in reference \cite{bal} that the twisted and untwisted ladder operators are related by a unitary transformation, thereby preserving the number operators}}
Since we always deal with states of fixed N, the operator $\hat N$ may be replaced by its eigenvalue N, thereby yielding a c-number contribution to the Hamiltonian
\begin{eqnarray}
\frac{N^{2}e^{2}}{V}\frac{2\pi}{\mu^{2}} - \frac{Ne^{2}}{V}\frac{2\pi}{\mu^{2}}.
\end{eqnarray}
The first term of the above expression cancels the first term of the Hamiltonian in (\ref{totham1}). The second term represents an energy $-2\pi e^{2}(V\mu^{2})^{-1}$ per particle and vanishes in the proper physical limit: first $L \rightarrow \infty$ and $\mu \rightarrow 0$ as discussed earlier. Thus the explicit $\mu^{-2}$ divergence cancells identically in the thermodynamic limit, which reflects the electrical neutrality of the total system; furthermore, it is now permissible to set $\mu = 0$ in the first term of (\ref{lastterm}), since the resulting expression is well defined. We therefore obtain the final Hamiltonian for a bulk electron gas in a uniform positive background
\begin{eqnarray}
\label{hamf}
\hat{H}= \sum_{{\bf k}\lambda}\frac{\hbar^{2}k^{2}}{2m}\tilde{a}^{\dag}_{{\bf k}\lambda}\tilde{a}_{{\bf k}\lambda} + \frac{2\pi e^{2}}{V}\sum_{\bf k k_{3} k_{4}}^{\prime} \sum_{\lambda_{1}\lambda_{2}}\frac{1}{k^{2}}e^{-i{\bf k}\wedge{\bf k}_{3}}e^{i{\bf k}\wedge{\bf k}_{4}}\tilde{a}^{\dag}_{{\bf k}+{\bf k}_{3},\lambda_{1}}\tilde{a}^{\dag}_{{\bf k}_{4}-{\bf k},\lambda_{2}}\tilde{a}_{{\bf k}_{4}\lambda_{2}}\tilde{a}_{{\bf k}_{3}\lambda_{1}}.
\end{eqnarray}
We can now introduce the inter-particle spacing $r_{0}$, defined through $V \equiv \frac{4}{3}\pi r_{0}^{3} N$ and measure it in units of the Bohr radius $a_{0} = \frac{\hbar^{2}}{me^{2}}$, thus yielding a dimensionless variable $r_{s}=\frac{r_{0}}{a_{0}}$

With $r_{0}$ as the unit of length we define the following quantities
\begin{eqnarray}
\bar V = r^{-3}_{0}V, \quad \quad    \bar {\bf k} = r_{0}{\bf k}, \quad \quad  \bar {\bf k}_{3} = r_{0}{\bf k}_{3},  \quad \quad  \bar {\bf k}_{4} = r_{0}{\bf k}_{4}
\end{eqnarray}
and thus we obtain the following dimensionless form of the Hamiltonian operator
\begin{eqnarray}
\hat{H} = \frac{e^{2}}{a_{0}r^{2}_{s}}(\sum_{\bar {\bf k}\lambda}\frac{1}{2}\bar {\bf k}^{2}\tilde{a}^{\dag}_{\bar {\bf k}\lambda}\tilde{a}_{\bar {\bf k}\lambda} + \frac{2\pi r_{s}}{\bar V}\sum^{\prime}_{[\bar{\bf k}\bar{\bf k}_{3}\bar{\bf  k_{4}}]}\sum_{\lambda_{1}\lambda_{2}}\frac{1}{\bar k^{2}} \tilde{a}^{\dag}_{\bar{\bf k} + \bar{\bf k}_{3},\lambda_{1}}\tilde{a}^{\dag}_{\bar{\bf k}_{4}-\bar {\bf k},\lambda_{2}}\tilde{a}_{\bar{\bf k}_{4}\lambda_{2}}\tilde{a}_{\bar{\bf k}_{3}\lambda_{1}}e^{-ir^{2}_{0}\bar{\bf k}\wedge(\bar{\bf k}_{3}-\bar{\bf k}_{4})}).
\end{eqnarray}
In the limit $r_{s} \rightarrow 0$, corresponding to the high density limit ($r_{0} \rightarrow 0$), the potential energy becomes a small perturbation even though it is neither weak nor short ranged, while the leading term comes from the kinetic-energy term. Thus the leading term in the interaction energy of a high-density electron gas can be obtained with first order perturbation theory. In the high-density limit, we can therefore separate the original dimensional form of the Hamiltonian (\ref{hamf}) into two parts:
\begin{eqnarray}
\hat{H}_{0} = \sum_{{\bf k}\lambda}\frac{\hbar^{2}k^{2}}{2m}\tilde{a}^{\dag}_{{\bf k}\lambda}\tilde{a}_{{\bf k}\lambda},
\end{eqnarray}
\begin{eqnarray}
\hat{H}_{1} = \frac{2\pi e^{2}}{V}\sum_{[{\bf k}{\bf k}_{3}{\bf k}_{4}]}^{\prime}\sum_{\lambda_{1}\lambda_{2}}\frac{1}{k^{2}}\tilde{a}^{\dag}_{{\bf k}+{\bf k}_{3},\lambda_{1}}\tilde{a}^{\dag}_{{\bf k}_{4}-{\bf k}, \lambda_{2}}\tilde{a}_{{\bf k}_{4}\lambda_{2}}\tilde{a}_{{\bf k}_{3}\lambda_{1}}e^{-i{\bf k}\wedge({\bf k}_{3}-{\bf k}_{4})},
\end{eqnarray}
where $\hat{H}_{0}$ is the unperturbed Hamiltonian, representing a noninteracting Fermi system, and $\hat{H}_{1}$ is the (small) perturbation. Correspondingly, the ground state energy $E$ may be written as $E^{0} + E^{1} + .....$, where $E^{0}$ is the ground-state energy of a free Fermi gas, while $E^{1}$ is the first-order energy shift. Since the Pauli exclusion principle allows only two fermions in each momentum eigenstate (hold even in noncommutative space){\footnote{As one can see that the Pauli's exclusion principle remains valid in the momentum space, as can be seen from (6) that in the case of $\eta \rightarrow -1$, $\hat{\hat{a}}_{{\bf k}}^{\dag}\hat{\tilde{a}}_{{\bf k}}^{\dag} = 0$. Consequently, the concepts like Fermi label etc. remains valid in momentum space even in presence of non-commutativity. However, this does not remain true in configuration space, as $\hat{\psi}^{\dag}({\bf x})\hat{\psi}^{\dag}({\bf x})$ does not vanish necessarily as follows from the relation $\psi(x)\psi(y)= \int{d^{2}x'd^{2}y'\Gamma_{\theta}(x, y, x', y')\psi(y')\psi(x')}; \theta \neq 0$\cite{chak}. Indeed it has been shown there, that repulsive statistical potential between a pair of identical (twisted) fermions can saturate to a finite value at coincident points, thereby violating Pauli principle in real space.}}, one with spin up and one with spin down, the normalized ground-state $|F>$ is obtained by filling the momentum states up to a maximum value, the Fermi momentum $p_{F} = \hbar k_{F}$. In the thermodynamic limit one can again replace summation by integration, so that $k_{F}$ can be determined by computing the expectation value of the number operator in the ground state $|F>$
\begin{eqnarray}
N = <F|\hat N|F> = \sum_{{\bf k} \lambda}<F|\hat n_{{\bf k}\lambda}|F> = \sum_{{\bf k}\lambda}\Theta (k_{F}-k) = (3\pi^{2})^{-1}Vk_{F}^{3} = N
\end{eqnarray}
where $\Theta (x)$ denotes the step function
\begin{eqnarray}
\Theta (x) = 1 \quad {\rm for} \quad x\geq 0, \quad
\Theta (x) = 0 \quad {\rm for} \quad x< 0.
\end{eqnarray}
Equivalently $k_{F}$ can be expressed as
\begin{eqnarray}
k_{F} = (\frac{3\pi^{2}N}{V})^{\frac{1}{3}} = (\frac{9\pi}{4})^{\frac{1}{3}} r_{0}^{-1}.
\end{eqnarray}
Now the expectation value of $\hat H_{0}$ may be evaluated as,
\begin{eqnarray}
E^{(0)} = <F|\hat{H}_{0}|F> = \frac{\hbar^{2}}{2m}\sum_{{\bf k}\lambda}k^{2}<F|\hat{n}_{{\bf k}\lambda}|F>
= \frac{\hbar^{2}}{2m}\sum_{{\bf k}\lambda}k^{2}\Theta(k_{F}-k)  \nonumber \\
             = \frac{\hbar^{2}}{2m}\sum_{\lambda}\frac{V}{(2\pi)^{3}}\int{d^{3}kk^{2}\Theta(k_{F}-k)}   
              = \frac{3}{5}\frac{\hbar^{2}k^{2}_{F}}{2m}N = \frac{e^{2}}{2a_{0}}\frac{N}{r_{s}^{2}}\frac{3}{5}(\frac{9\pi}{4})^{\frac{2}{3}}.
\end{eqnarray}
In a free Fermi gas, the ground-state energy per particle $E^{(0)}/N$ is $\frac{3}{5}$ of the Fermi energy $\epsilon_{F}^{0} = \hbar^{2}k_{F}^{2}/2m$.
We now compute the shift in the ground-state energy to first order in perturbation theory: 
\begin{eqnarray}
\label{pert}
E^{(1)} &=& <F|\hat{H}_{1}|F> \nonumber \\
        &=& \frac{2\pi e^{2}}{V}\sum_{{\bf k}{\bf k}_{3}{\bf k}_{4}}^{\prime}\sum_{\lambda_{1}\lambda_{2}}\frac{1}{k^{2}}<F|\tilde{a}^{\dag}_{{\bf k}+{\bf k}_{3},\lambda_{1}}\tilde{a}^{\dag}_{{\bf k}_{4}-{\bf k}, \lambda_{2}}\tilde{a}_{{\bf k}_{4}\lambda_{2}}\tilde{a}_{{\bf k}_{3}\lambda_{1}}e^{-i{\bf k}\wedge({\bf k}_{3}-{\bf k}_{4})}|F>.
\end{eqnarray}
For a non zero matrix element, all the states $|{\bf k}_{4}\lambda_{2}>$, $|{\bf k}_{3}\lambda_{1}>$, $|{\bf k}+{\bf k}_{3},\lambda_{1}>$, $|{\bf k}_{4}-{\bf k},\lambda_{2}>$ should be occupied in $|F>$.
Also, since ${\bf k}=0$ is excluded, we must have the pairing
\begin{eqnarray*}
{\bf k}+{\bf k}_{3}, \lambda_{1} \quad = \quad {\bf k}_{4}, \lambda_{2}
\end{eqnarray*}
so that the matrix element becomes
\begin{displaymath}
\delta_{{\bf k}+{\bf k}_{3},{\bf k}_{4}}\delta_{\lambda_{1}\lambda_{2}}<F|\tilde{a}^{\dag}_{{\bf k}+{\bf k}_{3},\lambda_{1}}\tilde{a}^{\dag}_{{\bf k}_{4}-{\bf k}, \lambda_{2}}\tilde{a}_{{\bf k}_{4}\lambda_{2}}\tilde{a}_{{\bf k}_{3}\lambda_{1}}|F>.
\end{displaymath}
The exponential factor in the previous expression for $E^{(1)}$ (\ref{pert}) again becomes the identity as ${\bf k}+{\bf k}_{3} = {\bf k}_{4} (\it i.e. \ {\bf k} = {\bf k}_{4}-{\bf k}_{3})$. Finally, using the twisted commutation relation given by (\ref{twcr1}), (\ref{twcr3}) and also considering the fact that the term ${\bf k} = 0$ is excluded from the sum, the above expression becomes
\begin{eqnarray}
-\delta_{{\bf k}+{\bf k}_{3},{\bf k}_{4}}\delta_{\lambda_{1}\lambda_{2}}<F|\hat{n}_{{\bf k}+{\bf k}_{3},\lambda_{1}}\hat{n}_{{\bf k}_{3}\lambda_{1}}e^{-2i{\bf k}_{3}\wedge({\bf k}+{\bf k}_{3})}|F>  \nonumber \\ = -\delta_{{\bf k}+{\bf k}_{3},{\bf k}_{4}}\delta_{\lambda_{1}\lambda_{2}}\Theta(k_{F}-|{\bf k}+{\bf k}_{3}|) \Theta(k_{F}-k_{3})e^{-2i{\bf k}_{3}\wedge{\bf k}}.
\end{eqnarray}
From this follows
\begin{displaymath}
E^{(1)}= -\frac{2\pi e^{2}}{V}\sum_{[{\bf k}{\bf k}_{3}\lambda_{1}]}\frac{1}{k^{2}}\Theta(k_{F}-|{\bf k}+{\bf k}_{3}|)\Theta(k_{F}-k_{3})e^{-2i{\bf k}_{3}\wedge{\bf k}}
\end{displaymath}
\begin{displaymath}
\quad \quad = -\frac{2\pi e^{2}}{V}\sum_{\lambda_{1}}(\frac{V}{(2\pi)^{3}})^2\int{d^{3}kd^{3}k_{3}\frac{1}{k^{2}}\Theta(k_{F}-|{\bf k}+{\bf k}_{3}|)\Theta(k_{F}-k_{3})e^{-2i{\bf k}_{3}\wedge{\bf k}}}.
\end{displaymath}
Note that we have again included the ${\bf k}=0$ term at the level of integration, as being a set of measure zero, it does not contribute anything to the integral.
Now taking the summation over $\lambda_{1}$ the above expression becomes
\begin{eqnarray*}
E^{(1)} = - \frac{4\pi e^{2}V}{(2\pi)^{6}}\int{d^{3}kd^{3}k_{3}k^{-2}\Theta(k_{F}-|{\bf k}+{\bf k}_{3}|)\Theta(k_{F}-k_{3})e^{-2i{\bf k}_{3}\wedge{\bf k}}}.
\end{eqnarray*}
It is convenient to change variables: ${\bf k}_{3} \rightarrow {\bf k}_{3} + \frac{\bf k}{2}$, which reduces the above expression into the symmetrical form
\begin{eqnarray}
E^{(1)} = - \frac{4\pi e^{2}V}{(2\pi)^{6}}\int{\frac{d^{3}k}{k^{2}}d^{3}k_{3}\Theta(k_{F}-|{\bf k}_{3}+\frac{1}{2}{\bf k}|)\Theta(k_{F}-|{\bf k}_{3}-\frac{1}{2}{\bf k}|)e^{-ik_{3i}\theta^{ij}k_{j}}}.
\end{eqnarray}
The computation of this integral can now be made simpler by orienting the 3rd
axis along the dual $ \thetav$ from eq.$(10)$, so that only $\theta^{12}\neq 0$
(see appendix A). Although a bit lengthy the integral can then be computed in a straightforward 
manner (see appendix B) to yield
\begin{eqnarray}
E^{(1)}= -\frac{16\pi^{3}Ve^{2}}{(2\pi)^{6}}k_{F}^{4}\sum_{r=0}^{\infty}\frac{(-1)^{r}(k_{F}^{2}\theta)^{2r}4^{-1-r}}{(1+r)(1+2r)(\Gamma(\frac{3}{2}+r))^{2}}.
\end{eqnarray}
This is a infinite series in $\theta$. Now as $\theta$ is of the order of Plank length-scale, hence very small, we sum the series up to the first contributing term in $\theta$, neglecting all other higher order terms. This gives
\begin{eqnarray}
E^{(1)} = -\frac{16\pi^{3}Ve^{2}}{(2\pi)^{6}}k_{F}^{4}(1-\frac{(k_{F}^{2}\theta)^{2}}{54}) = -\frac{e^{2}}{2a_{0}}[(\frac{9\pi}{4})^{\frac{1}{3}}\frac{3}{2\pi}\frac{1}{r_{s}} - (\frac{9\pi}{256})^{\frac{2}{3}}(\frac{\theta}{a_{0}^{2}})^{2}\frac{1}{r_{s}^{5}}].
\end{eqnarray} 
Thus the ground-state energy per particle in the high density limit is given approximately by
\begin{eqnarray}
\frac{E}{N} \quad  _{\buildrel{=}\over{r_{s}\rightarrow 0}} \quad \frac{e^{2}}{2a_{0}}[\frac{2.21}{r_{s}^{2}}-\frac{0.916}{r_{s}}+ (\frac{\theta}{a_{0}^{2}})^{2}\frac{0.2302}{r_{s}^{5}} + ......].
\end{eqnarray} 
As can be seen from the above expression, the ground-state energy has $\theta$ corrections and deviates from the commutative result. The ground-state energy per particle is increased by an order of ($\sim (\frac{\theta}{a_{0}^{2}})^{2}$), which is a dimensionless quantity. Note, however, that this $\theta$ dependency has its roots in the twisted anti-commutation relations, rather than the non-commutative structure of space per se, as the $\theta$ dependency arising from the star product, which encodes the non-commutativity of space, always dropped out. This is true to first order in perturbation theory.  However, as was shown in \cite{bem}) both these effects play a role to higher order in perturbation thery.  

It is also quite clear from the above expression that when $r_{s}$ becomes very small, $\it{i.e.}$ in the high-density limit, the effect of spatial non-commutativity on the ground-state energy of the degenerate electron gas becomes more significant. Taking the non-commutativity parameter $(\theta)$ to be of the order of plank-length it can be seen that the effect of non-commutativity in any terestrial experiments may not be found, but in the case of astrophysical objects, where matter density is very high, the effect of non-commutativity may be found by experiments.

%---------------------------------------------------------------------------
\section{Conclusions}
\label{con}
%---------------------------------------------------------------------------
We have shown that, in contrast to \cite{bem}, the ground-state energy of a non-commutative degenerate electron gas in a neutralising background acquires non-commutative corrections to first order in perturbation theory.  These corrections arise from the modified two particle correlations resulting from the twisted anti-commutation relations.  All $\theta$ dependency arising directly from the star product dropped out to this order in perturbation theory.  Our final observation is that any observable effect, as far as the energy shift is concerned, can only be obtained when the system is extremely dense - a situation that can presumably arise only in an astrophysical setting.

\section{Acknowledgement}

This work was supported by a grant under the Indo-South African research agreement
between Department of Science \& Technology, Government of India and South
African National Research Foundation. FGS would like to thank S.N.Bose National
Centre for Basic Sciences, Kolkata, for their hospitality during the time
when part of the work was completed. BC would like to thank the Institute
of Theoretical Physics, Stellenbosch University, for their hospitality during
the period of his stay there, when the investigation towards this problem
was initiated. BC would also like to thank T.Sreecharan for useful discussions.

%---------------------------------------------------------------------------
\section{Appendix A}
%---------------------------------------------------------------------------
Here we would like to demonstrate a problem in 3-dimension with spatial non commutativity can be reduced effectively to a problem of planar non commutativity by making an appropriate SO(3) transformation. 

To that end, let us first recall that the vector ${\thetav} = \{\theta_{i} \}$ dual to the non-commutative antisymmetric parameter $\theta^{ij}$ is defined as in eq.$(10)$,
\begin{eqnarray}
\theta_{i} = \frac{1}{2} \epsilon_{ijk} \theta^{jk} \nonumber.
\end{eqnarray}
Now this ${\thetav}$-vector will be pointed in an arbitrary direction. So we can parametrise it as
\begin{eqnarray}
{\thetav} = \theta\pmatrix{\sin\alpha \cos\beta \cr \sin\alpha \sin\beta \cr \cos\alpha} = \pmatrix{\theta^{23} \cr \theta^{31} \cr \theta^{12}}.
\end{eqnarray}
We would now like to rotate the coordinate axes $XYZ$ to $X'Y'Z'$ by making an SO(3) transformation in such a manner that $Z'$ axis now becomes parallel to the ${\thetav}$ direction. A convenient choice for the element of SO(3), doing this job, is obtained by taking products of two successive SO(2) rotation matrices around $Y$ and $Z$-axes as
\begin{eqnarray}
R = \left( \begin{array}{ccc} \cos\beta & -\sin\beta & 0 \\
                             \sin\beta & \cos\beta & 0 \\
                              0 & 0 & 1\end{array}\right)
\left( \begin{array}{ccc} \cos\alpha & 0 & \sin\alpha \\
                         0 & 1 & 0 \\
                         -\sin\alpha & 0 & \cos\alpha \end{array}\right) \nonumber \\ 
= \left( \begin{array}{ccc} \cos\alpha \cos\beta & -\sin\beta & \sin\alpha \cos\beta \\
                          \sin\beta \cos\beta & \cos\beta & \sin\alpha \sin\beta \\
                          -\sin\alpha & 0 & \cos\alpha \end{array}\right) \quad \in SO(3)
\end{eqnarray} 
As can be easily checked that the action of the $R$-matrix on the canonical orthonormal basis,
\begin{eqnarray}
{\bf e}_{1}=\pmatrix{1 \cr 0 \cr 0}; {\bf e}_{2}=\pmatrix{0 \cr 1 \cr 0}; {\bf e}_{3}=\pmatrix{0 \cr 0 \cr 1}; {\bf e}_{i}^{T}{\bf e}_{j} = \delta_{ij},
\end{eqnarray}
associated with the original $XYZ$-frame, yields following the orthonormal basis ${\bf e}_{i}^{\prime} = R {\bf e}_{i}$:
\begin{eqnarray}
{\bf e}_{1}^{\prime} = \pmatrix{\cos\alpha \cos\beta \cr \cos\alpha \sin\beta \cr -\sin\alpha}; {\bf e}_{2}^{\prime} = \pmatrix{-\sin\beta \cr \cos\beta \cr 0}; {\bf e}_{3}^{\prime} = \pmatrix{\sin\alpha \cos\beta \cr \sin\alpha \sin\beta \cr \cos\alpha}
\end{eqnarray}
for the $X'Y'Z'$-frame, as can read-off easily from the respective columns of the $R$-matrix (). Particularly, one can verify that ${\bf e}_{3}^{\prime}$ is indeed proportional to the ${\thetav}$-vector.

Now the position vector ${\bf r}$ can be expanded in either of these basis as, ${\bf r}=x^{i}{\bf e}_{i}={x^{\prime}}^{j}{\bf e}_{j}^{\prime}$.
Making use of the fact that ${\bf e}_{i}^{\prime T}{\bf e}_{j}^{\prime} = \delta_{ij}$, we can obtain ${x^{\prime}}^{i}$ in terms of $x^{j}$ as ${x^{\prime}}^{i} = x^{j}({{\bf e}^{i}}^{T}R^{-1}{\bf e}_{j})$. Writing elaborately, this yields
\begin{eqnarray}
\pmatrix{x' \cr y' \cr z'} = \left( \begin{array}{ccc} \cos\alpha \cos\beta & \cos\alpha \sin\beta & -\sin\alpha \\
                                                       -\sin\beta & \cos\alpha & 0 \\
                                                       \sin\alpha \cos\beta & \sin\alpha \sin\beta & \cos\alpha \end{array}\right) \pmatrix{x \cr y \cr z} \nonumber \\ 
= \pmatrix{ x\cos\alpha \cos\beta + y\cos\alpha \sin\beta -z\sin\alpha \cr -x\sin\beta + y\cos\beta \cr x\sin\alpha \cos\beta + y\sin\alpha \sin\beta + z\cos\alpha}.
\end{eqnarray}
Now, if the original $\hat{x}_{i}$'s are operator-valued and satisfy 
\begin{eqnarray}
[\hat{x}^{i}, \hat{x}^{j}] = i\theta^{ij}, \nonumber
\end{eqnarray}
then one can easily check that 
\begin{eqnarray}
[\hat{x}', \hat{y}'] = i\theta, \quad [\hat{y}', \hat{z}'] = [\hat{z}', \hat{x}']= 0. 
\end{eqnarray}
Thus the problem is essentially reduced to planner non-commutativity. 
This result is some-what expected, as the matrix $\{ \theta_{ij} \}$ is degenerate, as it is an odd dimensional $(3 \times 3)$ anti-symmetric matrix. And we have achived this by orienting the $z'$-axis of the rotated coordinate axes in the direction of ${\thetav}$-vector introduced in (10).

%---------------------------------------------------------------------------
\section{Appendix B}
%---------------------------------------------------------------------------
Here we provide some of the basic steps leading to the evaluation of the integral
\begin{eqnarray}
I = \int{\frac{d^{3}k}{k^{2}}d^{3}k_{3}\Theta(k_{F}-|{\bf k}_{3}+\frac{1}{2}{\bf k}|)\Theta(k_{F}-|{\bf k}_{3}-\frac{1}{2}{\bf k}|)e^{-ik_{3i}\theta^{ij}k_{j}}}.
\end{eqnarray}
These $\Theta$ functions indicate that the exponential function has to be integrated first over the overlapping region '$R$' of the two spheres of equal radii $k_{F}$ centred at $\pm(\frac{1}{2}{\bf k})$ with respect to the variable ${\bf k}_{3}$ holding ${\bf k}$ fixed. The resultant function of ${\bf k}$ is then integrated over the solid sphere $|{\bf k}| \le 2k_{F}$. Considering the region $R_{+}(k_{3z}>0)$ initially, we can write the corresponding integral over $k_{3}$ with proper limit as,
\begin{displaymath}
L_{+} = \int_{R_{+}}{d^{3}k_{3}e^{-ik_{3i}\alpha^{i}}} \quad \quad \quad {\rm where} \quad \quad \alpha^{i}= \theta^{ij}k_{j}
\end{displaymath}
\begin{displaymath}
\quad = \int_{0}^{(k_{F}-\frac{k}{2})}{dz e^{-i\alpha^{3}z}}\int_{-\sqrt{(k_{F}^{2}-(z+\frac{k}{2})^{2})}}^{\sqrt{(k_{F}^{2}-(z+\frac{k}{2})^{2})}}{dy e^{-i\alpha^{2}y}}\int_{-\sqrt{(k_{F}^{2}-(z+\frac{k}{2})^{2}-y^{2})}}^{\sqrt{(k_{F}^{2}-(z+\frac{k}{2})^{2}-y^{2})}} {dx e^{-i\alpha^{1}x}}.
\end{displaymath}
Here we have substituted $k_{3x}=x$, $k_{3y}=y$ and $k_{3z}=z$ for convenience.

Upon simplification, this takes the following form,
\begin{displaymath}
L_{+}= \frac{2}{\alpha^{1}}\int_{0}^{(k_{F}-\frac{k}{2})}{dz e^{-i\alpha^{3}z}}\int_{-p}^{+p}{dye^{-i\alpha^{2}y}\sin(\alpha^{1}\sqrt{p^{2}-y^{2}})}  \quad {\rm where} \quad p=\sqrt{k_{F}^{2}-(z+\frac{k}{2})^{2}}. 
\end{displaymath}

Substituting the above expression for $L_{+}$ in (62), one gets for the corresponding $I_{+}$
\begin{displaymath}
I_{+} \equiv \int{\frac{d^{3}k}{k^{2}}L_{+}} = \int{\frac{d^{3}k}{k^{2}}} [\frac{4}{\alpha^{1}}\int_{0}^{(k_{F}-\frac{k}{2})}{dz}\int_{0}^{p}{dy Cos(\alpha^{2}y)Sin(\alpha^{1}\sqrt{p^{2}-y^{2}})}], 
\end{displaymath}
where we have taken $\theta^{13}= \theta^{23} = 0$, by orienting the 3-axis in ${\bf k}$ frame in the direction of ${\thetav}$, where $\theta_{i}= \frac{1}{2}\epsilon_{ijk}\theta_{jk}$ (see (10) and Appendix A) so that the only surviving componenets of $\theta^{ij}$ are $\theta^{12}= -\theta^{21} = \theta$ consequently $\alpha^{3}=0$ and $\alpha^{1}=\theta k_{2}$ and $\alpha^{2} = -\theta k_{1}$.
The above integral can now be expressed in terms of Bessel functions \cite{grad}
\begin{displaymath}
J_{\nu}(z)= \frac{z^{\nu}}{2^{\nu}}\sum_{r=0}^{\infty}(-1)^{r}\frac{z^{2r}}{2^{2r}r!\Gamma(\nu+r+1)}
\end{displaymath}
as
\begin{displaymath}
I_{+}= 2\pi \int{\frac{d^{3}k}{k^{2}}}\int_{0}^{(k_{F}-\frac{k}{2})}{dz\frac{p}{\sqrt{(\alpha^{1})^{2}+(\alpha^{2})^{2}}}J_{1}(p\sqrt{(\alpha^{1})^{2}+(\alpha^{2})^{2}})}.
\end{displaymath}

Now we can integrate term by term by expanding series of Bessel functions. By making use of the integral
\begin{displaymath}
\int_{0}^{(k_{F}-\frac{k}{2})}{dz(k_{F}^{2}-(z+\frac{k}{2})^{2})^{r+1}} = (k_{F})^{(2r+3)}\sum_{j=0}^{r+1}\frac{^{r+1}C_{j}(-1)^{j}}{2j+1}(1-(\frac{k}{2k_{F}})^{(2j+1)}),
\end{displaymath}
one gets after a lengthy but straight forward computation
\begin{displaymath}
I_{+}= -k_{F}^{4}\sum_{r=0}^{\infty}\frac{(-1)^{r}(k_{F}^{2}\theta)^{2r}}{r!(r+1)!(2r+1)!}\sum_{j=0}^{r+1}\frac{^{r+1}C_{j}(-1)^{j}}{(r+j+1)}\sum_{l=0}^{r}\frac{^{r}C_{l}(-1)^{l}}{(2l+1)}.
\end{displaymath}
The last two factors involving series summation can now be carried out exactly to get
\begin{displaymath}
\sum_{l=0}^{r}\frac{^{r}C_{l}(-1)^{l}}{(2l+1)} = \frac{\sqrt{\pi}r!}{2\Gamma(\frac{3}{2}+r)},
\end{displaymath}
\begin{displaymath}
\sum_{j=0}^{r+1}\frac{^{r+1}C_{j}(-1)^{j}}{(r+j+1)} = \frac{4^{(-1-r)}\sqrt{\pi}r!}{\Gamma(\frac{3}{2}+r)}.
\end{displaymath}
With this $I_{+}$ takes the following form
\begin{eqnarray}
I_{+} = 4\pi^{2}k_{F}^{4}\sum_{r=0}^{\infty}\frac{(-1)^{r}(k_{F}^{2}\theta)^{2r}4^{-1-r}}{2(1+r)(1+2r)(\Gamma(\frac{3}{2}+r))^{2}}.
\end{eqnarray}
A similar result is obtained for $I_{-}$, when one considers the region $R_{-}$ corresponding to $k_{3z}<0$. Consequently one has $I= 2I_{+}=2I_{-}$.
Substituting this back in (54), one gets the desired expression of $E^{(1)}$.

\end{document}